\documentstyle[aps,prl,multicol,epsf]{revtex}
\renewcommand{\narrowtext}{\begin{multicols}{2} \global\columnwidth20.5pc}
\renewcommand{\widetext}{\end{multicols} \global\columnwidth42.5pc}
\renewcommand{\v}[1]{{\bf #1}}

\newcommand{\s}{{\sigma}}

\newcommand{\gr}{{\nabla}}
\def\eqa{\begin{eqnarray}}
\def\eea{\end{eqnarray}}
\newcommand{\eq}{\begin{equation}}
\newcommand{\ee}{\end{equation}}
\newcommand{\nn}{\nonumber\\}
\newcommand{\Eq}[1]{Eq.~(\ref{#1})}
\newcommand{\p}{\partial}

\newcommand{\cL}{ {\cal L} }

\begin{document}
\draft
\title{Antiferromagnetism, Stripes, and Superconductivity in the\\ 
t-J Model with Coulomb Interaction}
\author{Jung Hoon Han$^{(a)}$, Qiang-Hua Wang$^{(a),(b)}$ and Dung-Hai Lee$^{(a)}$}

\address{${(a)}$Department of Physics,University of California
at Berkeley, Berkeley, CA 94720, USA \\}
\address{${(b)}$ Physics Department and National Laboratory of Solid State Microstructures,\\ Institute for Solid State Physics, Nanjing University, Nanjing 210093, China\\}

\maketitle
\draft
\begin{abstract}
We study mean-field phases of the t-J model with long-range Coulomb interaction. In the order of increasing doping density we find a classical antiferromagnet, charge and spin stripes, and a uniform $d$-wave superconductor, at
the realistic doping parameters. Both in-phase and anti-phase stripes exist as metastable configurations, but the
in-phase stripes have a slightly lower energy. The dependence of the stripe width and the inter-stripe spacing on
the doping is examined. Effects of fluctuations around the mean-field states are discussed.
\end{abstract}

\narrowtext 
The cuprate materials which exhibit high-T$_c$ superconductivity show many different ordering
tendencies as the hole doping concentration ($x$) in the system is varied. At zero doping the cuprates are
antiferromagnetic insulators below the N\'{e}el temperature (Fig. 1(a))\cite{AFM}. The ordered antiferromagnetic phase
ceases to exist with more than about $2\%$ of holes. Around $5\%$ doping the system begins to show
superconductivity at low temperatures. In the intervening $2-5$\% the system is lacking either antiferromagnetic
order or superconductivity. In the doping range between $5\%$ and $15\%$, superconductivity co-exists with the
(dynamic) stripe order\cite{stripereview,INS,zhou}. The stripes are most readily seen in neutron scattering experiments at
non-zero energy transfer, hence ``dynamic", where one finds evidence of one-dimensional periodic modulation of the
antiferromagnetic order as well as the charge density\cite{INS}.

As noted from the early days of high-T$_c$, the cuprates are characterized by strong electron-electron
repulsion\cite{anderson}, which gives rise to a Mott-insulating state at half-filling. The t-J model, which
incorporates such electron repulson, has been extensively studied in connection with high-T$_c$ superconductivity.
Although it is quite likely that the t-J model correctly captures the short-distance correlation of electrons in
the cuprates, one must not {\it a priori} overlook the long-range part of Coulomb interaction. In particular for
the t-J model without Coulomb interaction, phase separation occurs for a wide range of doping
concentration\cite{hellberg}, which pre-empts the possibility of a high-pairing-scale superconducting state.

Various mean-field theories which {\it assumes} a uniform ground state have been in existence\cite{mf}. These theories have varying degrees of success in understanding the phases of the cuprates.
Such mean-field approaches have however been subject to the skepticism that the no-double-occupancy constraint is
treated only approximately. Attempts to improve the treatment of the constraint result in a strongly fluctuating gauge field.
Recently one of us (D.-H.L.) was able to integrate out the gauge field exactly at low energies in the uniform superconducting phase\cite{lee}.
It was shown that despite the drastic modification of the excitations, mean-field vacua serves as a good starting
point in characterizing the zero-temperature state.

This work is performed under the assumption that mean-field theory will capture the short-distance/high-energy
ordering tendency of the t-J + Coulomb model, and that the long-distance/low-energy properties can be understood by studying the soft fluctuations of the mean-field order parameters.

The Hamiltonian we consider is the following: \eqa H&=&-t\sum_{\langle ij\rangle} (b_jb^{\dag}_i f_{j\alpha}^\dag
f_{i\alpha} \!+h.c.) +J\sum_{\langle ij \rangle} (\v S_i\cdot\v S_j \! -\frac{1}{4}\!n_{i}n_{j})\nonumber \\
 &+& {V_c\over 2}\sum_{i\ne j} {1\over {r_{ij}}}(n_{i}-\bar{n})(n_{j}-\bar{n}).
\label{h} \eea The no-double-occupancy constraint is expressed as $f^{\dag}_{i\alpha} f_{i\alpha} +b^{\dag}_i b_i
-1=0$ in terms of the spinon ($f_{i\sigma}$) and holon ($b_i$) operators. Other notations include
$n_{i}=1-b^{\dag}_ib_i$ (site electron density), and $\bar{n}=1-x$ (average electron density). Our first goal is
to understand the mean-field phases sustained by this model. Unlike other mean-field theories in the past\cite{mf,sachdev}, we include the 
magnetic order parameter on an equal footing with all other order parameters. Among other things this gives us the possibility of obtaining the long-range ordered antiferromagnetic state observed in experiments.
\\

\noindent
{\bf Mean-field theory}

Our mean-field theory, in essence, is a variational approach: a trial wavefunction is constructed and parameters
are varied to obtain the minimum energy. We construct a wavefunction which allows local magnetic moments (but not
limited to antiferromagnetism), superconducting pairing (but not limited to $d$-wave symmetry), and modulations in
the charge density. The trial wavefunction $|\Psi \rangle$ is given by $|\Psi \rangle=|\Psi_b \rangle \otimes
|\Psi_f \rangle$ where the bosonic $(|\Psi_b \rangle)$ and fermionic ($|\Psi_f \rangle)$ states are independently
constructed from their respective vacua  as follows: 
\eqa &&|\Psi_b \rangle =(\chi_j b^\dag_j)^{N_b}|0_b
\rangle\nn &&|\Psi_f \rangle=\prod_{a,b} \left(u^{a}_jf^\dag_{j\uparrow}+v^{a}_jf_{j\downarrow}\right)
\left(w^{b}_lf^\dag_{l\downarrow}+z^{b}_lf_{l\uparrow}\right) |0_f \rangle. \label{ans} 
\eea 
Repeated indices $j$
and $l$ implies summation over lattice sites. The bosons are assumed to be condensed, and the fermion ground state is
constructed by occupying the (yet undetermined) quasiparticle orbitals labeled by $a,b$.  The mean-field
single-particle wavefunctions $u^a_j, v^a_j, w^b_l, z^b_l$ and $\chi_j$ are varied to minimize 
\eqa
 \langle\Psi|H|\Psi \rangle&-&\sum_i\lambda_i \langle\Psi|b^{\dag}_ib_i+
f^{\dag}_{i\alpha}f_{i\alpha}-1|\Psi \rangle\nn 
&-&\mu\sum_i\langle \Psi| n_i -\bar{n} |\Psi \rangle .
\label{qu}
\eea
Lagrange multipliers $\lambda_i$, and $\mu$ are introduced to guarantee that the {\it average occupation} obeys the constraints locally as well as globally.

The calculation is carried out numerically on a $N_x \times N_y$ lattice with a periodic boundary condition. Not assuming any translational invariance, we first perform  totally unrestricted
minimization for  $N_x, N_y\le 16$. 
After the nature of the solution is established, we perform a more restricted search for $N_x, N_y$ up to $120$. The results reported below are for $t/J=3$, $V_c /J=5$. Other choices of $t/J$ and $V_c/J$ are also
studied, with results that are not qualitatively different from those presented below. 

We  find three prominent types of order. In the order of increasing doping, they are antiferromagnetic insulator, charge/spin stripes, and uniform $d$-wave superconductor.

\widetext
\begin{figure}[ht]
\hskip -0.5cm \centering \epsfxsize=17cm \epsfysize=3cm \epsfbox{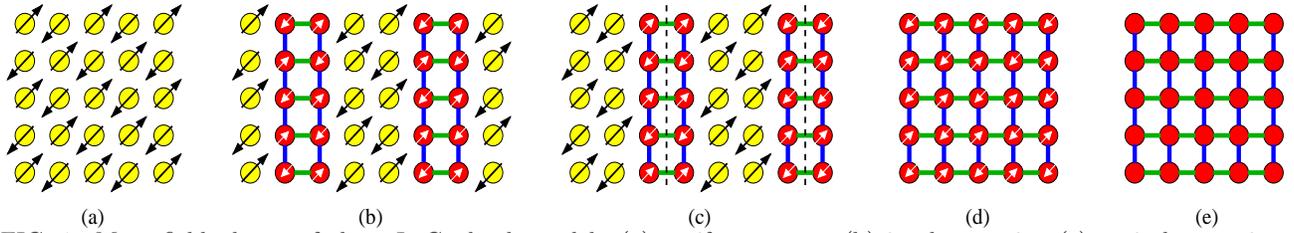} \caption{Mean-field phases of the
t-J+Coulomb model: (a) antiferromagnet  (b) in-phase stripe (c)
anti-phase stripe (d) uniform $d$-wave superconductor with residual antiferromagnetic order (e) non-magnetic uniform $d$-wave
superconductor. Yellow/red circle represents sites fully/partially occupied 
by the electrons. Magnetic moments for partially occupied sites  (white arrow)  are smaller than those of  fully occupied sites (black arrow). The bonds
shown in black and green are the pairing amplitudes, which differ in sign for $x$ and $y$ directions. The dashed vertical lines in Fig. 1(c) indicate the boundary where the shift of the antiferromagnetic order parameter occurs.}
\label{fig:den}
\end{figure}
\narrowtext

{\it Antiferromagnet (Fig. 1(a))}: At zero doping the mean-field ground state shows antiferromagnetic long-range
order. Each electron is surrounded by four neighbors with opposite spins. It is an insulator because the occupation
constraint forbids the electrons to hop. This mean-field state is a caricature of the insulating
antiferromagnet observed in the undoped cuprates\cite{AFM}. For $x\!<\!x_{c1}\!\approx\! 0.02$, the doped
holes are localized, often in the form of elongated puddles (or finite-length stripes). However, these localized
puddles do not disrupt the overall antiferromagnetic order.

{\it Stripes (Fig. 1(b)-(c))}: For $x_{c1}\!<\!x\!<\!x_{c}\approx 0.14$ the mean-field ground state shows charge
corrugation in the form of stripes. A stripe is a region which is extended in one direction (say $\hat{y}$) and
localized in the other, with partially occupied sites. There are two types of stripes, anti-phase and in-phase.
The antiferromagnetic order parameter goes through a $\pi$ phase shift across the anti-phase stripe, whereas it
remains in-phase across the in-phase stripe. As a result the anti-phase stripe modulates the antiferromagnetic
order with a period twice that for the charge density.
\begin{figure}[ht]
\hskip  -0.2cm \centering \epsfxsize=4cm \epsfysize=3.5cm
\epsfbox{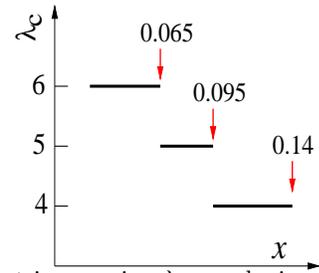} \caption{The stripe spacing
$\lambda_c$ vs. doping for the in-phase stripes. The numbers above the arrow indicate the approximate doping fraction where the periodicity changes.}
\end{figure}

We find that the {\it in-phase} stripes are the ground states for 
$x_{c1}\!<\! x \!<\! x_{c}$. At $x=x_c$ a first
order phase transition occurs, after which the system becomes uniform. In all cases we have studied the in-phase
stripes are bond-centered and have a width of two lattice constants. The stripe spacing, on the other hand,
depends on doping and increases as the doping level decreases. The smallest distance between the in-phase stripes
we observe is four lattice constants, and it occurs near $x_c$. This spacing is maintained in the doping range $
0.095\le x\le 0.14$. For this range the site hole density in each stripe varies from $0.19$ per site
(corresponding to a line density of $0.38$) at $x=0.095$ to $0.28$ (line density $0.56$) at $x=0.14$. As the doping
decreases below $0.095$ a new stripe configuration with the stripe spacing equal to five lattice constants emerges
as the ground state. The stripe width is still two. The evolution of the stripe spacing with the doping
concentration is shown in Fig. 2. We infer from this the existence of a series of integer stripe spacings as
doping decreases. Each spacing has a non-zero range of stability giving rise to plateaus in the modulation period
of the charge density. The step-wise evolution of the stripe spacing is clearly a lattice commensuration effect. In the presence of thermal fluctuation of the shape of stripes (quantum fluctuation is
known not to roughen the stripe shape), we expect a smoother evolution of the modulation period. Inside each stripe there exists strong superconducting pairing as well as weak
magnetism, as illustrated in Fig. 1(b). The pairing gap inside the in-phase stripe is comparable to the maximal pairing gap ($\approx 0.05J$) observed in the uniform $d$-wave phase, while the magnetic moments are a fraction of the full moment, $\langle S_z \rangle =\pm 1/2$, of the insulator. An in-phase stripe is in some sense an optimum pairing state which is confined in the one-dimensional geometry.  As in-phase stripes get closely
spaced, it is likely that transverse fluctuation smears out the charge corrugation and results in a high-pairing-scale superconductor.

In the entire range of $x_{c1}\le x\le x_c$ the anti-phase stripes are metastable mean-field solutions. The energy
difference between the most favorable anti-phase and in-phase stripes is shown as function of doping in Fig. 3. The largest difference ($10\%$) occurs at $x=0.025$ and the smallest ($0.14\%$) at $x_c$.
Note that anti-phase stripes come very close in energy to the
in-phase ones near $x_c$, and therefore, fluctuations that are omitted by the mean-field theory may stabilize the anti-phase stripes. (One such candidate is the transverse fluctuation of the stripes.)  When that happens, the progression of the ground states vs. doping will be according to Figs. 1(a) through 1(e). We  discuss the properties of the anti-phase stripe in the following.

The anti-phase stripes are also bond-centered and have  a width equal to two lattice constants. The stripe spacing evolves in a step-wise fashion similar to Fig. 2. The range of hole density inside the anti-phase stripe is consequently similar to the in-phase stripe case. The anti-phase stripes also have non-zero pairing and magnetic moments (Fig. 2(c)). However, 
the pairing scale is considerably smaller (by a factor of three) than that in the in-phase stripes. In this sense, {\it anti-phase stripes are unfavorable as far as pairing is concerned}. 

{\it Uniform $d$-wave superconductor (Fig. 1(d)-(e))}: The homogeneous phase, $x>x_c$, is characterized by
$d$-wave pairing and, for $x$ near $x_{c}$, some residual antiferromagnetism\cite{comment}. The pairing scale is maximum at $x=x_c$ where $\Delta_{max}\approx 0.05 J$ and decreases monotonically as $x$ increases.  The antiferromagnetic
moments disappear completely for $x\ge 0.2$. 
\begin{figure}[ht]
\hskip -0.2cm \centering \epsfxsize=4.5cm \epsfysize=3.5cm
\epsfbox{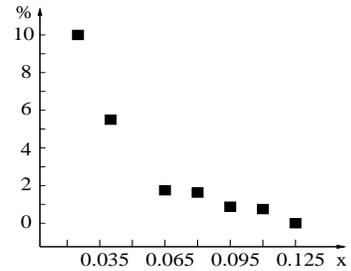}
\caption{Energy difference of the anti-phase and the in-phase stripes, 
$2(E_{anti}\!-\!E_{in})/(E_{anti}\!+\!E_{in})$, as a function of $x$. In-phase stripe has an increasingly lower energy as the average hole density diminishes.}
\end{figure}
We find it gratifying that our mean-field theory produces states which extrapolate between  extreme classical (antiferromagnet at $x=0$) and  extreme quantum (uniform superconducting) limits. 

{\it Coulomb interaction and high pairing scale:} In light of the above mean-field results, we argue that {\it
high-$T_c$ superconductivity is a cooperative effect due to the short-range magnetic exchange and the long-range
Coulomb interaction.}\cite{kivem}  Larger antiferromagnetic exchange favors higher pairing scale, however it also causes phase
separation to set in at a lower doping level and the uniform high-pairing state becomes inaccesssible. 
At $t/J=3$, our model shows a phase-separated ground state for $x\le
0.26$ in the absence of Coulomb interaction. Coulomb interaction suppresses phase separation and produces two
compromises -- stripes and a high-pairing-scale $d_{x^2-y^2}$ superconductor. 
\\

\noindent{\bf Fluctuations}

The low-energy excitations often appear in the form of the fluctuation of the
order parameters. In the present context these include:  phase fluctuation ($\theta_b$) of the Bose
condensate $|\Psi_b\rangle$, phase fluctuation ($\theta_p$) in the pairing condensate of $|\Psi_f \rangle$,
orientation fluctuation of the magnetic moments ($\hat{\Omega}$), shape fluctuation and displacement of the
stripes, gapless (neutral) fermion excitation in the case of $d$-wave pairing, and most importantly the ``gauge
fluctuation'' inherent in the slave-boson theory\cite{gauge}.

Due to the occupancy constraint and the form of \Eq{h} there exists an internal gauge symmetry\cite{gauge} under a
local phase change, 
$
b_i\rightarrow e^{i\theta_i}b_i, ~f_{i\alpha}\rightarrow e^{i\theta_i}f_{i\alpha}.
$
Such
symmetry is broken by most of the mean-field vacua, giving rise to a fluctuating gauge field as the soft mode.
Recently one of us were able to integrate out the gauge fluctuations exactly in the non-magnetic, uniform $d$-wave
superconducting phase corresponding to Fig. 1(e)\cite{lee}. The result is the confinement of two Goldstone modes,
$\theta_b$ and $\theta_p$, into one $\phi=2\theta_b-\theta_p$,  which is the phase of the electron superconducting
condensate. The final low-energy dynamics is that of a phase-fluctuating superconductor with gapless fermion
excitations\cite{BFN,lee}: $\cL=\cL_{\phi}+\cL_{\psi}+\cL_{int}$, where 
\eqa
&&\cL_{\phi}=\frac{K}{2}(\gr\phi)^2+\frac{1}{2u}(\p_t\phi)^2 +i\bar{\rho}\p_0\phi\nn &&\cL_{\psi}=\sum_{n=1}^2
\bar{\psi}_{n\alpha}(\p_t-iv_{xn}\tau_x\p_x-iv_{yn}\tau_z\p_y)\psi_{n\alpha} \nn
&&\cL_{int}=-iz_{\mu}j_{\psi\mu}\p_{\mu}\phi. 
\label{efa} 
\eea 
In the above $\bar{\rho}$ is the average Cooper pair density, $\psi_n$ is the fermion field associated with the $n$th gap node, $\tau_z$ is the third Pauli matrix, $v_{xn},v_{yn}$ specify the linear
dispersion of the gapless fermions,  and
$j_{\psi\mu}=\frac{1}{2}(\sum_n\bar{\psi}_{n\s}\tau_z\psi_{n\s} ,iv_{x1}\bar{\psi}_{1\s}\psi_{1\s},
iv_{y2}\bar{\psi}_{2\s}\psi_{2\s})$, is the fermion 3-current. Due to the gauge fluctuation the parameters $K,u,z_{\mu}$ are strongly renormalized. In particular $K$ is proportional to $x$,  which accounts for the low superfluid density in spite of the high pairing scale in the pseudogap regime. 
 Similar treatment of gauge fluctuations has been
done for each of the mean-field phases discussed above. The results are somewhat technical and will be reported
elsewhere.

For the antiferromagnet the mean-field state satisfies the occupation constraint exactly. In this phase the only
low-energy degree of freedom is the fluctuation of the direction of the local spin. The spin degrees of freedom is
gauge-neutral and hence unaffected by strong gauge fluctuations. The interaction between the spin waves is
described by the familiar non-linear sigma model\cite{CHN} \eq \cL_{\Omega} =\frac{K_{\sigma}}{2}\left[
\frac{1}{v^2}|\p_t\hat{\Omega}|^2+|\gr\hat{\Omega}|^2\right]. 
\label{nls} 
\ee 
In two space dimensions, it is well
known that the spin-wave fluctuation does not destroy the antiferromagnetic long-range order as long as the spin
stiffness $K_{\sigma}$ is sufficiently big. This certainly seems to be the case for the undoped
cuprates\cite{AFM}.

In the uniform $d$-wave phase corresponding to Fig. 1(d) there exists residual antiferromagnetic moments. The
low-energy degrees of freedom are those of the non-magnetic superconductor plus the spin fluctuation. Since the
wavevector associated with the commensurate antiferromagnetic ordering, $\v k=(\pi,\pi)$, mismatches the momentum
connecting the gap nodes, the magnetic degrees of freedom decouple from the low-energy fermions. The effective
action is then simply the sum of \Eq{efa} and \Eq{nls}. Due to the smallness of the magnetic moments in this phase,
however, the quantum fluctuation is likely to wash out the long-range correlation of the residual magnetism. In
that case, the distinction between magnetic and non-magnetic superconductors becomes obscure.

The fluctuations in the stripe phase is the richest. The low-energy degrees of freedom include the phase of the
superconducting condensate and fermion quasiparticle excitations inside the stripes, the fluctuation of the
magnetic moment, and the displacement and shape fluctuation of the stripes. Unlike the antiferromagnetic and the
uniform superconducting phases, the fluctuation of stripes can not only modify the properties of a given stripe
phase, but also change the energy ordering between the in-phase and anti-phase stripes. Despite some
progress\cite{KFE}, a complete picture which
involves all of the above fluctuations is still lacking. The subject  is currently under investigation. 

{\bf Acknowledgment}\rm$~$
We are indebted to Steve Kivelson for numerous helpful discussions. We also thank Eduardo Fradkin, Z.-X. Shen and Ned Wingreen for insightful remarks and questions. Part of the numerical calculations are carried out with the computing facility at NEC research. DHL is supported by NSF grant DMR 99-71503. QHW is supported by the National Natural Science Foundation of China, the National Centre for Research and Development of Superconductivity, and the Berkeley Scholars Program.

\widetext
\end{document}